\documentclass[12pt]{article}

\input amssym.def
\input amssym
\newcommand{\be}{\begin{equation}}
\newcommand{\ee}{\end{equation}}

\title{Li\'enard--Wiechert solution revisited}
\author{Nikolai V. Mitskievich\thanks{Physics Department,
CUCEI, University of Guadalajara, Guadalajara, Jalisco, Mexico.}
\thanks{Postal address: Apartado Postal 1-2011, C.P. 44100,
Guadalajara, Jalisco, M\'exico. E-mail:
mitskievich03@yahoo.com.mx}}
\date{~}

\begin{document}
\renewcommand{\theequation}{\arabic{section}.\arabic{equation}}
\maketitle

\begin{abstract}
A self-sufficient consideration of the Li\'enard--Wiechert
solution is given including its heuristic deduction, which
involves a future light cone (thus with lightlike propagation of
information from an arbitrarily moving pointlike charge), and
physical interpretation of this field {\it via} application of
three distinct reference frames (of an inertial observer, then a
non-inertial one retardedly co-moving with the charged source, and
finally, co-moving with the electromagnetic field). In the last
frame the magnetic part of the Li\'enard--Wiechert field
identically (though not asymptotically) vanishes in all spacetime
together with the Poynting vector. In the second frame, the
properties of energy redistribution and radiation are discussed.
The dynamically caused propagation velocity of the
Li\'enard--Wiechert electromagnetic field in a vacuum at any final
distance from the source is less than that of light.
\end{abstract}

~~~~~~~~~~~\\

Propagation of electromagnetic field. Classification of
electromagnetic fields. Transforming away the magnetic field.
Reference frame co-moving with the electromagnetic field.

\newpage

\section{Introduction} \label{s1}

\subsection{Preliminaries} \label{s1.1}

In this paper we consider Maxwellian electromagnetic fields in the
flat Min\-kowski spacetime with the metric tensor $g_{\mu\nu}=
\textnormal{diag}(+1,-1,-1,-1)$, thus taking Cartesian coordinates
(algebraic relations used or deduced here, frequently remain
unaltered also in the framework of general relativity). Greek
indices are four- and Latin, three-dimensional. However we more
frequently use as three-dimensional quantities four-dimensional
vectors (tensors) orthogonal to the timelike unit vector
describing the reference frame (the monad). Different frames may
be simultaneously applied (the test-object property which is
essential in treatment of reference frames in non-quantum theory).
In general, we do not mutually relate reference frames and systems
of coordinates. A comma ($\,_,$) followed by an index is used to
denote partial differentiation with respect to the corresponding
coordinate. We also use natural units in which the velocity of
light in a vacuum is $c=1$. Round brackets mean symmetrization and
square brackets, antisymmetrization in the indices contained in
them (the so-called Bach brackets). In concrete calculations no
approximations are assumed.

This material is essentially the final chapter of my unpublished
one-semester course ``Relativistic Physics'' given to
undergraduate (Licenciatura) students at the Physics Department of
the University of Guadalajara during the last seven years. The
students first have to attend another course on tensor calculus
which also includes the formalism of Cartan forms with some
applications in physics. The subsection \ref{s4.2} was included in
my course only in the last semester.

\subsection{Preview of the paper} \label{s1.2}

More than one hundred years ago, A. Li\'enard \cite{Lien} and E.
Wiechert \cite{Wiech} discovered an exact solution of Maxwell's
equations describing electromagnetic field of a pointlike electric
charge in an arbitrary motion. A frequently used treatment of this
solution can be found in \cite{LanLif}, and its more general
deduction, including the use of an arbitrary mixture of retarded
and advanced potentials, in \cite{SyngeSR}. In section~\ref{s2} we
consider a simple and direct deduction of the Li\'enard--Wiechert
(below abbreviated as LW) solution with the use of the light cone
concept which involves a supposition of lightlike propagation of
information from this pointlike source. Some important general
properties of the LW solution are discussed in section~\ref{s3}.
Here a general classification of electromagnetic fields is
outlined, and it is found that the LW field belongs to the pure
electric type, thus its magnetic part can be transformed away when
one passes to certain non-inertial reference frames. It is well
known that in a vacuum electromagnetic waves propagate with the
fundamental velocity $c$ ($=1$). However, as it is shown in
section~\ref{s4}, a mixture of non-radiative and radiative
electromagnetic fields has another propagation velocity ($<1$).
For this reason, when we speak above and in sections \ref{s2} and
\ref{s5} about `propagation of information,' we do not speak
strictly about propagation of electromagnetic field in the general
sense. In subsection~\ref{s4.2} the general method of finding
reference frames co-moving with electromagnetic fields is
formulated (mostly for the case of pure subtypes of electric or
magnetic types of fields {\it via} transformation away of the
magnetic or electric field, respectively; however also in the
impure subtypes, though there it is impossible to transform away
one or --- asymptotically
--- both fields {\bf E} and {\bf B}, one always may make these
fields mutually parallel, thus transforming away the Poynting
vector in the respective frame). In frames co-moving with the
electromagnetic field, the Poynting vector automatically vanishes.
This method is then applied to the LW field. Relative motion of
different reference frames is considered in subsection~\ref{s4.3},
first in general and then for the LW solution. In section~\ref{s5}
some results obtained in the paper are discussed. In two
appendices, \ref{sA} and \ref{sB}, a short review of the
Ehlers--Zel'manov covariant theory of reference frames (its
algebraic part) is given together with applications to the
description of electric and magnetic fields.

\section{A systematic deduction of the LW
solution} \setcounter{equation}{0} \label{s2}

Let us consider a pointlike charge $Q$ in a motion along a
worldline $L$ parametrically described as \be \textnormal{{\bf
r}}'=\textnormal{{\bf r}}'(t'), ~ \mathrm{ equivalently, } ~
x'^i=x'^i(t'), \ee $t'=x'^0$, $i=1, ~ 2, ~ 3$. We shall determine
at an arbitrary, but fixed spacetime point $P$ with coordinates
$x^\mu$ (not on $L$), the electromagnetic field created by the
charge $Q$ being at another point $P'$ on $L$; the coordinates are
chosen to be Cartesian. It is obvious that the electromagnetic
field created by a pointlike charge should have a singularity on
$L$, this is why we exclude here the case of coincidence of the
points $P$ and $P'$. Note that the coordinates of $P$ represent
four independent scalar variables $x^\mu$, and those of $P'$
merely are scalar functions of some parameter (this may be $s'$,
but we shall use the retarded time $t'$) along the worldline $L$,
$x'^\mu(t')$ (there are three equations, the fourth being simply
an identity, $x'^0=t'$). To mutually relate the spacetime points
$P$ and $P'$, we use a hypothesis that the information about
position and state of motion of the charge propagates with the
fundamental velocity (that of light) in an accordance with the
relativistic causality law. If the point $P$ and worldline $L$ are
given, the point $P'$ can be determined as that of intersection of
the past light cone with a vertex at $P$ and the line $L$ (this
simultaneously means that $P$ is on the future light cone with a
vertex at $P'$). This constructive definition is important in the
subsequent calculations, but fortunately the concrete relation
between the position of $P$ and the corresponding retarded time
$t'$ at $P'$ turns out to be of no importance. Thus $t'$ is a
function of all four coordinates of $P$ --- we write it as
$t'(x)$; we shall easily calculate the explicit form of
derivatives of $t'$ with respect to the coordinates $x^\mu$
without an explicit knowledge of $t'(x)$.

We take the Minkowski metric as $g_{\mu\nu}=g^{\mu\nu}=
\textnormal{diag} (+1,-1,-1,-1)$ (in fact, this is the definition
of Cartesian coordinates), thus the tangent vector to $L$,
$u'^\mu=dx'^\mu/ds'$ (the four-velocity of the charge) taken at
the retarded point $P'$, is timelike and unitary ($u'\cdot u'
\equiv u'^\mu u'_\mu=1$), its timelike property being manifested
by the relation $ds'^2>0$ along $L$. Locally, $u'$ determines the
direction of growth of the proper time $s'$, being simultaneously
the projector onto the (retarded) physical time direction of the
(retarded) reference frame (retardedly) co-moving with the charge.
Another projector, now a tensor, can be constructed as
(\ref{bproj}), here \be \label{bmunu} b_{\mu\nu}=
g_{\mu\nu}-u'_\mu u'_\nu. \ee It is (a) symmetric ($b_{\mu\nu}=
b_{\nu\mu}$), (b) orthogonal to $u'$, thus realizing projection
onto the subspace $\perp u'$ (the physical three-space of the just
mentioned inertial reference frame at $P'$); (c) it possesses the
property of idempotent ($b^\mu_\lambda b^\lambda_\nu=b^\mu_\nu$
with det\,$b^\mu_\lambda= 0$), and (d) plays the r\^ole of the
three-dimensional metric in the mentioned subspace, with the
signature $(0,-,-,-)$ (zero is inserted in the four-dimensional
sense). Thus $g^\lambda_\lambda \equiv\delta^\lambda_\lambda=4$
and $b^\lambda_\lambda=3$ give dimensionalities of the space-time
and subspace under consideration.

Let us introduce a vector connecting the four-points (events) $P'$
and $P$, \be \label{R} R^\mu=x^\mu-x'^\mu(t'). \ee Of course, this
is not a vector under more general transformations than the
Lorentz ones (like the Euclidean `radius vector' is a vector only
in Cartesian systems). Since $R^\mu$ lies on the light cone, \be
\label{RR} R^\mu R_\mu=0, \ee this vector is null. Its projection
onto $u'$ is denoted as $D$, and onto the retarded three-space, as
$\textnormal{\bf D}^\mu$: \be \label{D} D:=u'^\mu R_\mu\equiv
u'\cdot R, ~ ~ ~ \textnormal{\bf D}^\mu=R^\nu b^\mu_\nu=R^\mu-
Du'^\mu, ~ ~ \textnormal{\bf D}\perp u'. \ee Due to (\ref{bmunu}),
$\Rightarrow \delta^\mu_\nu=b^\mu_\nu+u'^\mu u'_\nu$, and the null
property (\ref{RR}), \be \label{D.D} \textnormal{\bf D}^\mu
\textnormal{\bf D}_\mu=-D^2, ~ ~ ~ D=\sqrt{-\textnormal{\bf
D}^\mu\textnormal{\bf D}_\mu}, \ee thus we call $\textnormal{\bf
D}^\mu$ the `retarded spatially projected vector between $P'$ and
$P$.' Similarly, $D$ is interpreted as the retarded
three-dimensional distance between $P'$ and $P$. Recall also that
\be \label{uexp} u'^\mu=\frac{dx'^\mu}{ds'}=
\frac{dx'^0}{ds'}\frac{dx'^\mu}{dx'^0}=u'^0\cdot(1,v'^i). \ee

Now we are ready to calculate all necessary derivatives (of $t'$,
$R^\mu$, $D$, $u'^\mu$, and more) with respect to $x^\mu$. The
first step is to write
$$
{R^\mu}_{,\alpha}=\frac{\partial x^\mu}{\partial
x^\alpha}-\frac{\partial x'^\mu}{\partial
x^\alpha}=\delta^\mu_\alpha-\frac{dx'^\mu}{ds'}\frac{ds'}{dt'}
\frac{\partial t'}{\partial x^\alpha},
$$
that is, \be \label{t,} {R^\mu}_{,\alpha}=\delta^\mu_\alpha-
\frac{u'^\mu}{u'^0} t'_{,\alpha}. \ee Differentiation of
(\ref{RR}) yields
$$
R_\mu{R^\mu}_{,\alpha}\equiv\frac{1}{2}\left(R_\mu R^\mu
\right)_{,\alpha}=0,
$$
thus \be t'_{,\alpha}=\frac{u'^0 R_\alpha}{D}, \ee and its
substitution into (\ref{t,}) yields \be
{R^\mu}_{,\alpha}=\delta^\mu_\alpha-\frac{u'^\mu R_\alpha}{D}. \ee
Now, \be \label{diffu} {u'^\mu}_{,\alpha}=\frac{du'^\mu}{dt'}
t'_{,\alpha}= \frac{du'^\mu}{ds'}\frac{ds'}{dt'}t'_{,\alpha}=
\frac{a'^\mu R_\alpha}{D} \ee (similar derivatives of all primed
objects are proportional to $R$ with the differentiation
subindex), where \be a'^\mu= \frac{du'^\mu}{ds'} \ee is the
acceleration four-vector (at $P'$) obviously possessing the
property of four-orthogonality to $u'$: \be u'^\mu a'_\mu\equiv 0.
\ee This use of the acceleration four-vector is more economic than
of the respective three-vector, though their mutual relation is
somewhat indirect; the reader, beginning with (\ref{a4-3}), may
easily reconstruct the corresponding formulae and apply them to
interpretation of the results and to make a comparison with the
treatment of LW problem in \cite{LanLif}. The final step in this
part of calculations is to differentiate $D$: \be D_{,\alpha}=(u'
\cdot R)_{,\alpha}=u'_{\mu, \alpha}R^\mu+u'_\mu{R^\mu}_{,\alpha}
=u'_\alpha-\frac{R_\alpha}{D}(1-a'\cdot R) \ee where, of course,
$a'\cdot R:= a'_\mu R^\mu\equiv a'\cdot\textnormal{\bf D}$. Let us
also take into account that \be \label{divR} {R^\nu}_{,\nu}=3 ~
\textnormal{ and } {a'^\mu}_{,\nu}= \frac{da'^\mu}{ds'}\,\!
\frac{R_\nu}{D} \ee [see a comment to (\ref{diffu})].

The second, and last, preparatory part of our calculations is to
write down Maxwell's equations. Outside the sources, their
four-dimensional form is \be \label{Max} {F^{\mu\nu}}_{,\nu}=0 \ee
where \be F_{\mu\nu}=A_{\nu,\mu}-A_{\mu,\nu} \ee is the field
tensor written in terms of the four-potential $A_\mu$, thus
${F^{\mu\nu}}_{,\nu}=\square A^\mu+\left({A^\nu}_{,\nu}\right)^{,
\mu}=0$, the d'Alembertian operator being $\square=\Delta-
\partial^2/\partial t^2$. The ${A^\nu}_{,\nu}$-term can be
eliminated if we use the Lorenz condition\footnote{This condition
is due not to H.A.~Lorentz as admits the majority of physicists,
but to L.V.~Lorenz (born in Elsinore, Denmark, in 1829), see the
footnote related to formula (5.1.47) in \cite{PenRin}, p. 321.}
\be \label{LorCond} {A^\nu}_{,\nu}=0 \ee which only fixes global
gauge of the four-potential without any other restrictions. The
alternative form of Maxwell's equations should then include the
Lorenz condition, thus in the form of a system \be \label{Maxw}
\square A^\mu=0 ~ ~ \textnormal{and} ~ ~ {A^\nu}_{,\nu}=0. \ee The
well-known Coulomb potential in a vacuum in electrostatics can be
written as $A^\mu=\frac{Q}{r}\delta^\mu_0$ for a pointlike charge
$Q$ located at the spatial origin. One notices that the
four-velocity of the charge at rest is $u'^\mu=u^\mu=
\delta^\mu_0$. This potential exactly satisfies both equations of
(\ref{Maxw}) when $\textnormal{{\bf r}}\neq 0$. We shall now show
that a simple generalization of the Coulomb potential is also an
exact solution of Maxwell's equations, and this is precisely that
of Li\'enard--Wiechert.

The generalization is simply \be \label{LW} A^\mu=
\frac{Qu'^\mu}{D}. \ee The proof that this is the exact solution
is quite short for the Lorenz condition:
$$
{A^\nu}_{,\nu}=\frac{Q}{D}\left({u'^\nu}_{,\nu}-\frac{u'^\nu
D_{,\nu}}{D}\right)=\frac{Q}{D^2}\left[a'\cdot R-1+\frac{R_\nu
u'^\nu}{D}\left(1-a'\cdot R\right)\right]\equiv 0,
$$
and for the d'Alembert equation [the first in (\ref{Maxw})], a
little tedious. First, we calculate \be \label{Amunu} A_{\mu,
\nu}=\frac{Q}{D^2}\left[a'_\mu R_\nu-u'_\mu\left(u'_\nu
-R_\nu\frac{1-a'\cdot R}{D}\right)\right]. \ee Turning now to the
rest of (\ref{Maxw}), we see that it is necessary to consider
$\square A_\mu=-{A_{\mu,\nu}}^{,\nu}$, taking into account
(\ref{divR}) and the already known derivatives of $u'$, $a'$,
$R^\alpha$, and $D$. The reader can verify after performing
differentiation that for $D\neq 0$ all terms identically cancel:
$$
\left\{\frac{Q}{D^2}\left[a'_\mu R^\nu-u'_\mu\left(u'^\nu-R^\nu
\frac{1-a'\cdot R}{D}\right)\right]\right\}_{,\nu} \equiv 0.
$$
This completes the proof.

Since we shall need the full expression of $F_{\mu\nu}$ in the
subsequent calculations, let us now antisymmetrize the expression
(\ref{Amunu}) (the first term in round brackets is immediately
cancelled): \be \label{Fmunu} F_{\mu\nu}=\frac{Q}{D^2}\left[R_\mu
\left( a'_\nu+u'_\nu \frac{1-a'\cdot R}{D}\right)- R_\nu\left(
a'_\mu+u'_\mu \frac{1-a'\cdot R}{D}\right) \right]. \ee This is a
specific type of skew-symmetric tensor sometimes called {\it
simple bivector} since it represents an antisymmetrization of only
two vectors, $R^\mu$ (\ref{R}) and $U^\mu=\frac{Q}{D^2}\left(
a'^\mu+u'^\mu \frac{1-a'\cdot R}{D} \right)$: \be \label{FRU}
F_{\mu\nu}=R_\mu U_\nu-U_\mu R_\nu \ee which can be written as a
2-form $F=R\wedge U$, $R=R_\mu dx^\mu$ and $U=U_\mu dx^\mu$.

\section{General properties of the LW field}
\setcounter{equation}{0} \label{s3}

First it is worth mentioning the obvious fact that the Coulomb
field is a special case of the LW solution: one simply has to
consider a pointlike charge at rest, that is $u'^\mu
=\delta^\mu_0$ for any $P'$, thus $a'^\mu=0$. This is the reason
why the LW solution has to be interpreted as the electromagnetic
field of an arbitrarily moving pointlike charge (of course, the
Gauss theorem is here also applicable, for example, in an inertial
frame instantaneously co-moving with the central charge at $P'$).

\subsection{Classification of electromagnetic fields and its
application to the LW solution} \label{s3.1}

The classification of electromagnetic fields is based on existence
of only two invariants built with the field tensor $F_{\mu\nu}$,
while all other invariants are merely algebraic functions of these
two (if not vanish identically). The first invariant is
$I_1=F_{\mu\nu} F^{\mu\nu}=2(\textnormal{{\bf
B}}^2-\textnormal{{\bf E}}^2)$, and the second, $I_2=
F\!\stackrel{\textnormal{\small$\ast$}}{
\textnormal{\scriptsize$\mu\nu$}} F^{\mu\nu}=4 \textnormal{{\bf
E}}\bullet\textnormal{{\bf B}},$ {\it cf.} (\ref{elE}) and
(\ref{magB}); the definition of $I_2$ contains dual conjugation of
$F_{\mu\nu}$, \be \label{Fdual}
F\!\stackrel{\textnormal{\small$\ast$}}{
\textnormal{\scriptsize$\mu\nu$}}:=
\frac{1}{2}\epsilon_{\mu\nu\alpha\beta}F^{\alpha\beta}, ~ ~
F\!\stackrel{\textnormal{\scriptsize$\mu\nu$}}{
\textnormal{\small$\ast$}}:=-
\frac{1}{2}\epsilon_{\mu\nu\alpha\beta}F_{\alpha\beta}. \ee Here
$\epsilon_{\mu\nu\alpha\beta}$ is the completely skew-symmetric
object (not exactly a tensor) with $\epsilon_{0123}=+1$, known as
the Levi-Civit\`a symbol. In fact, only the squared $I_2$ is
really invariant, and $I_2$ itself is a pseudo-invariant which
acquires the factor $J/|J|$ by a general transformation of
coordinates, $J$ being the Jacobian of the transformation, thus
the concrete sign of $I_2$ does not matter. In terms of $I_1$ the
invariant classification suggests three types of fields: $I_1<0$
is the electric type (the electric field dominates), $I_1>0$ gives
the magnetic type, and to $I_1=0$ corresponds the null type. On
the pseudo-invariant $I_2$ the further working out in detail of
the classification is based: the additional subtypes are impure
($I_2\neq 0$) and pure ($I_2=0$). It is important that the pure
electric case permits (at least, locally, if one considers only
inertial frames) to completely eliminate the magnetic field, and
similarly, the pure magnetic field permits to completely eliminate
the electric field, while the pure null electromagnetic field in a
vacuum permits to find a coordinate system (reference frame) in
which the electric and magnetic field intensities would take any
desired finite (nonzero and non-infinite) and equal values, but,
of course, the field will continue to pertain to the same pure
null type (in this case, both fields {\bf E} and {\bf B} will be
ever equal in their absolute values and mutually orthogonal, as
can be seen from the structure of both invariants). This last
property is closely related to the Doppler effect (not only in the
sense of the frequency, but --- and more profoundly --- also of
the field intensity), in particular, a complete elimination of the
pure null type field is `possible' only asymptotically (in less
rich-in-content terms, this means `impossible'), since there
cannot exist any reference frame moving with the speed of light
with respect to an arbitrary permissible reference frame. The
impure electric, magnetic, and null types obviously do not permit
such manipulations with the three-dimensional parts {\bf E} and
{\bf B} of the electromagnetic field (in the impure electric and
magnetic cases it is impossible to transform away the counterparts
of these respective fields).

Let us now apply this classification to the LW electromagnetic
field. Since $I_2=\frac{1}{2}\epsilon_{\mu\nu
\alpha\beta}F^{\mu\nu} F^{\alpha\beta}\equiv 0$ for any simple
bivector (\ref{FRU}), even with {\it arbitrary} $R$ and $U$, the
field is pure. Then it is pure electric since \be \label{I1} I_1=
-\frac{2 Q^2}{D^4}<0 \ee (remarkably, the structure of $I_1$ is
exactly Coulombian). This means that at any point of the spacetime
(any finite value of the distance $D$, {\it i.e.} not
asymptotically) it is possible to transform away the magnetic part
of the field; moreover, it is possible to find such a global
reference frame in which only electric part of the field will be
present. This possibility can be globally realized for any
concrete choice of the motion of the pointlike charge. In these
specific reference frames which are in general non-inertial, but
naturally admissible in special relativity (like those to which we
are accustomed in non-relativistic physics, the area much more
restricted than special relativity), the Poynting vector of the LW
field will vanish globally. This fact will be discussed in more
concrete details below. Its physical meaning is that at any finite
point of the spacetime the electromagnetic LW field propagates
with sub-luminal velocity.

\section{Propagation of the LW electromagnetic
field} \setcounter{equation}{0} \label{s4}

\subsection{Viewpoint of an inertial observer} \label{s3.2}

This is the least interesting case of the reference frame
application to LW solution while the approach reduces to use of a
monad adapted to Cartesian coordinates. Let the inertial observer
at $P$ measure electric and magnetic fields {\bf E} and {\bf B} as
well as electromagnetic energy density $w$ and Poynting vector
{\bf S} which are two of the three decomposition parts of
(\ref{Tmunu}) (we shall not consider the stress tensor) with
respect to this observer's monad $\tau^\mu=\delta^\mu_0$ (the
observer is at rest with respect to the Cartesian coordinates) and
to the corresponding orthogonal projector
$b^\mu_\nu=\delta^\mu_\nu- \delta^\mu_0\delta_\nu^0
\Leftrightarrow \delta^\mu_i\delta_\nu^j \delta^i_j \Rightarrow
(0,\delta^i_j)$ [see (\ref{bmunu})], \be \label{winert} w\equiv
{T_{ \textrm{\scriptsize em}}}^0_0= \frac{1}{4\pi}\left(
\frac{1}{4}F_{\alpha\beta}F^{\alpha\beta}-
F_{0\alpha}F^{0\alpha}\right)= \frac{1}{8\pi}\left(
\textnormal{{\bf E}}^2+\textnormal{{\bf B}}^2\right), \ee \be
\label{Poynting} \textnormal{{\bf S}}^i=
{T_{\textnormal{\scriptsize em}}}^i_0=-\frac{1}{4\pi}F_{i\alpha}
F^{0\alpha}=\frac{1}{4\pi}\left(\textnormal{{\bf E}}\times
\textnormal{{\bf B}}\right)^i, \ee {\it cf.} (\ref{Ttau}). Here
\be \label{EBinert} \textnormal{\bf E}^i=\frac{Q}{D^3}\left[u'_0
\left(R^i-R_0v'^i \right)(1-a'\cdot R)+D\left(a'_0R^i-R_0a'^i
\right) \right] \ee and $\textnormal{\bf B}^i=\ast(dt\wedge
R\wedge U)\equiv(\textnormal{\bf n}\times \textnormal{\bf E})^i$
where (for the inertial frame) $\textnormal{\bf n}=\textnormal{\bf
R}/R_0$ and the electromagnetic field 2-form $F=R \wedge U$ where
$R$ and $U$ are 1-forms built of the respective covectors found in
(\ref{FRU}); see also the definitions (\ref{EB}) and (\ref{magB}).
Taking into account (\ref{a4-3}) and relations
$D=u'_0\left(R_0-R^iv'^i\right)$ and
$R^i\left(R^i-R_0v'^i\right)=R_0\left(R_0-R^iv'^i\right)$, it is
easy to verify that (\ref{EBinert}) coincides with the expression
given by Landau and Lifshitz (\cite{LanLif}, (63,8)) --- in our
notations, \be \label{LLE} \textnormal{\bf E}^i=\frac{Q}{\left(
R_0-R^iv'^i \right)^3}\left\{\frac{1}{{u'_0}^2}
\left(R^i-R_0v'^i\right)+\left[ \textnormal{\bf
R}\times\left((\textnormal{\bf R}-R_0\textnormal{\bf
v}')\times\dot{\textnormal{\bf v}}'\right) \right]\right\}. \ee
However, since the Poynting vector expression is nonlinear in
characteristics of the electromagnetic field (due to
multiplication of electric and magnetic vectors), we prefer our
consideration given in the next subsection to that which splits
(\ref{LLE}) in two parts one of which should describe the outgoing
radiation; this reasoning works only asymptotically, and the
expression (\ref{EBinert}) is in this case more transparent than
(\ref{LLE}) due to the factor $D$ in the corresponding term in
square brackets in (\ref{EBinert}).

Finally, it is worth mentioning that the differential
characteristics of any inertial frame (its acceleration, rotation,
and rate-of-strain tensor), including the frame considered above,
identically vanish, thus of course simplifying the considerations
given in \cite{LanLif}, though at the cost of omission of some
important details.

\subsection{The retarded reference frame co-moving with the
charge} \label{s4.1}

The retarded reference frame at $P$ co-moving with the charge at
$P'$ is determined by the monad $\tau^\mu=u'^\mu$. Thus for
electric and magnetic fields we have \be \label{BEretard}
\textnormal{{\bf E}}=\frac{Q}{D^2}\left[(1-a'\cdot
R)\textnormal{{\bf n}}-D\textnormal{{\bf a}}'\right], ~ ~
\textnormal{\bf B}=\frac{Q}{D}\,\textnormal{\bf a}'\!\times\!
\textnormal{\bf n}. \ee However it is more direct to use
projections considered in appendix \ref{sB} which result in the
non-inertial reference frame where the Poynting vector is
$$
S^\mu:=T^\nu_\lambda u'^\lambda b^\mu_\nu,
$$
and projections have to be applied to $F_{\lambda\alpha}
F^{\nu\alpha}$ using (\ref{FRU}) and the relation $R_\mu U^\mu=
Q/D^2$ obvious from $U^\mu$ given just before that expression for
$F_{\mu\nu}$. Then
$$
F_{\lambda\alpha}F^{\nu\alpha} u'^\lambda b^\mu_\nu=
\frac{Q^2}{D^4}\left[D\,\textnormal{\bf D}^\mu\, a'\!\cdot\!a'-D\,
a'^\mu-\frac{\textnormal{\bf D}^\mu}{D}\,a'\!\cdot\!
\textnormal{\bf D}(1-a'\!\cdot\!\textnormal{\bf D})\right].
$$
In order to find a more concise form of the last expression, let
us introduce the unit radial vector $n$ perpendicular to the
monad: \be \label{nproj} \textnormal{\bf n}^\sigma:=
\frac{\textnormal{\bf D}^\sigma}{D}, ~ ~ ~ \textnormal{\bf
n}\cdot\textnormal{\bf n}=-1. \ee Then
$$
S^\mu=-\frac{Q^2}{4\pi D^2}\left[\textnormal{\bf n}^\mu\left(a'\!
\cdot\! a'+(\textnormal{\bf n}\!\cdot\! a')^2\right)-\frac{1}{D}
\left( a'^\mu+\textnormal{\bf n}^\mu(\textnormal{\bf n}\!\cdot\!
a')\right)\right].
$$
This expression however takes more transparent form if we also use
the projector onto the two-dimensional surface simultaneously
orthogonal to both $u'$ and $n$. This will be a spherical surface
of radius $D$ not in a hyperplane perpendicular to $u'$, but on
the future light cone with its vertex at $P'$ (a sphere
corresponding to the retarded time in analogy with determination
of the LW field). Thus we introduce the projector \be
\label{cproj} c^{\sigma\tau}:= b^{\sigma\tau}+n^\sigma
n^\tau\equiv \eta^{\sigma\tau}-u'^\sigma u'^\tau+n^\sigma n^\tau,
\ee
$$
c^{\sigma\tau}c_{\rho\tau}=c^\sigma_\rho, ~ ~ c^{\sigma\tau}
n_\sigma=0, ~ ~ {c^\sigma}_\sigma=2,
$$
and relations similar to $a^\epsilon\equiv b^{\epsilon\sigma}
a_\sigma$ should be also taken into account. This new projection
tensor plays the r\^ole of metric tensor on the two-dimensional
sphere with the signature $(0,0,-,-)$ involving two zeros, one
with respect to direction of the proper time from the viewpoint of
all four dimensions, and the second, in the sense of the radial
direction ($n$) which corresponds to the sphere. Finally, the
Poynting vector takes the form \be \label{simplPoynt} S^\mu=
\frac{Q^2}{4\pi}\left(\frac{1}{D^3} c^{\mu\tau} a'_\tau-
\frac{1}{D^2}n^\mu c^{\sigma\tau}a'_\sigma a'_\tau\right). \ee

This remarkably simple expression of the LW energy flux suggests
the following two conclusions. First, the part proportional to
$1/D^3$ and linear in the retarded four-acceleration $a'$, is
perpendicular to the radial direction {\bf n} ({\it i.e.}, it is
restricted to the corresponding two-sphere on the future light
cone with its vertex at $P'$). Thus it describes a redistribution
of energy at the fixed retarded distance $D$ from the field
source. The integral redistribution flux becomes smaller with more
distant location of the observer and asymptotically ($D
\rightarrow\infty$) tends to zero due to multiplication of
(\ref{simplPoynt}) by the two-dimensional surface element of the
sphere ($\sim D^2$), while the integration is performed only in
the sense of angular coordinates on the sphere. Of course, the
very surface (if taken not on the light cone), as well as the
reference frame's three-space, is non-holonom since in general
this frame possesses rotation \be \label{rotu'} \omega=\ast(u'
\wedge du')=\ast(a'\wedge u'\wedge n)= \textnormal{{\bf a}}'\times
\textnormal{{\bf n}}, \ee see (\ref{rot}), (\ref{vecprod}),
(\ref{diffu}), and the final remarks in appendix \ref{sA}. Second,
the part proportional to $1/D^2$ has positive radial direction
(take into account that it gives exactly this contribution since
four-dimensional square of the spacelike vector $a'$ is negative
due to the space-time signature). Thus it describes an energy flux
{\it from} the charge {\it to} spatial infinity. Moreover, all
this part of energy really goes to infinity without being
accumulated or rarefied at any values of $D$. Hence this term
really describes radiation of energy by the accelerated charge.
The non-holonomicity remark is here also relevant, and in this
situation one has to take certain caution; this is why we
mentioned a roundabout approach involving the light cone which
always exists and represents a real hypersurface, though its
normal vector is null, thus at the same time it is on the light
cone itself. This problem goes beyond the bounds of our paper, and
we only mention here that it was successfully treated in last few
decades in general relativity. After all, we are living and
working in the rotating reference frame of our planet, therefore
our three-dimensional physical space certainly is non-holonom, but
this does not prevent us to do physics and to apply it quite well.

In the retarded co-moving reference frame of the pointlike charge
the LW electromagnetic energy flux has no other constituent parts.
Since the problem does not take into account the sources of
acceleration of the charge (the lack of a strict auto-consistency
of the problem), the energy flux does not result here in any
change of the state of motion of charge. One may say that there is
implicitly some kind of engine which prescribes the exact world
line of the charged particle (the LW problem does not involve any
information about the particle's mass and energy), thus this
``engine'' automatically ``takes into account'' the particle's
energy loss due to radiation (which at finite distances is not
ligtlike, see below). Other details follow from the further
consideration of a new reference frame in which the magnetic field
of the LW solution simply vanishes.

\subsection{LW solution in the
reference frame co-moving with electromagnetic field, but not with
the charge} \label{s4.2}

In a reference frame which is co-moving with electromagnetic
field, the Poynting vector should vanish. This can occur for two
alternative reasons (to be realized in this frame): either
electric and magnetic vectors are mutually parallel (this is the
impure classification subcase), or one of them is equal to zero
(the pure subcase). The first case was considered by Wheeler
\cite{Whee} toward other ends. The second case pertains naturally
to the LW field since this is a pure electric one (thus Wheeler's
approach is not applicable, and the magnetic part can be
transformed away {\it via} a proper choice of the reference
frame). In fact, this possibility is scarcely encountered in
literature (I even don't know any references), and it would be
interesting to investigate it in more detail. We shall see that
this task is much simpler than one could expect.

Remember the general form of the LW field tensor, (\ref{FRU}):
$F_{\mu\nu}=R_\mu U_\nu-U_\mu R_\nu$. Let us (algebraically)
regauge the vector $U ~ \rightarrow ~ V=U+kR$ where $k$ is a
scalar function. This does not change the field tensor, \be
\label{FRV} F_{\mu \nu}=R_\mu V_\nu-V_\mu R_\nu. \ee Applying now
the 1-form definition of the magnetic vector in a $\tau$-frame
(\ref{magB}) and taking the monad as $\tau=NV$ where the scalar
normalization factor is $N=(V\cdot V)^{-1/2}$, we obviously come
to {\bf B}$=0$ in this frame. The problem is thus reduced to a
proper choice of $k$ such that $V$ will be a suitable real
timelike vector with $V\cdot V>0$. This method should work in our
case (for a pure magnetic field, a similar technique can be
applied, though requiring automatic representation of $\ast F$ as
a simple bivector).

We see that \be \label{V1} V^\mu=\frac{Q}{D^2}\left(
a'^\mu+\frac{1-a'\!\cdot\! R}{D}\,u'^\mu+kR^\mu \right), \ee thus
it was natural to include before $k$ the scalar coefficient
$Q/D^2$. Then \be V\!\cdot\! V=\left(\frac{Q}{D^2}\right)^2\left[
a'\!\cdot\! a'+ \frac{(1-a'\!\cdot\! R)^2}{D^2}+2k\right]. \ee In
fact, $k$ still remains arbitrary. Let it be \be k=
\frac{1}{2}\left[\frac{1}{D^2}-a'\!\cdot\! a'-\frac{(1-a'\!\cdot\!
R)^2}{D^2}\right] \ee (the first term in the square brackets,
$1/D^2$, got its denominator to fit the dimensional
considerations). Finally, \be V\cdot V=\left(\frac{Q}{D^3}
\right)^2>0 \ee and \be \label{taucomov} \hat{\tau}^\mu=Da'^\mu+
\left(1-a'\!\cdot\! R\right)u'^\mu+\frac{1}{2D}\left[1-D^2a'\!
\cdot\! a'-\left(1-a'\!\cdot\! R\right)^2\right]R^\mu \ee (it is
clear that $\hat{\tau}\cdot\hat{\tau}=+1$). By its definition, the
monad $\hat{\tau}$ describes the reference frame co-moving with
the LW electromagnetic field: in this frame the Poynting vector of
the field vanishes, and the electromagnetic energy flux ceases to
exist due to the absence of magnetic part $\hat{\textnormal{\bf
B}}$ of the field in this frame (applicable at any finite distance
$D$, not asymptotically). Really, (\ref{FRV}) now can be rewritten
as
$$
F_{\mu\nu}=\frac{Q}{D^3}\left(R_\mu\hat{\tau}_\nu-\hat{\tau}_\mu
R_\nu\right),
$$
thus the expression of $\hat{\textnormal{\bf B}}$ (\ref{magB})
contains $\hat{\tau}\wedge R \wedge\hat{\tau}\equiv 0$.

Let us now calculate the electric vector $\hat{\textnormal{\bf
E}}$ in the frame $\hat{\tau}$. A combination of (\ref{taucomov}),
(\ref{V1}), and (\ref{FRV}) gives \be \label{FRVCart} F=R\wedge
V=\frac{Q}{D^3}R \wedge\hat{\tau}, \ee see also (\ref{F2form}).
Then the expression (\ref{elE}) yields \be \label{Ecomov}
\hat{\textnormal{\bf E}}=\ast(\hat{\tau} \wedge\ast F)=
\frac{Q}{D^3}\ast[\hat{\tau}\wedge\ast(R\wedge\hat{\tau})]=
\frac{Q}{D^2}\hat{\textnormal{{\bf n}}} \ee which is, up to an
understandable reinterpretation of notations, exactly the form
known as the Coulomb field vector. Here $\hat{\textnormal{\bf
n}}^\mu=\hat{\textnormal{\bf D}}^\mu/D$ ($\perp\hat{\tau}$) where
$R^\mu u'_\mu=: D\equiv\hat{D}:=R^\mu\hat{\tau}_\mu$ and
$\hat{\textnormal{\bf D}}^\mu\!\!=\hat{b}^\mu_\nu R^\nu$ with
$\hat{b}^\mu_\nu=\delta^\mu_\nu- \hat{\tau}^\mu\hat{\tau}_\nu$,
hence \be \label{Dhat} \hat{\textnormal{\bf D}}^\mu=-D^2a'^\mu-
D\left(1-a'\! \cdot\! R\right)u'^\mu+\frac{1}{2}\left[1+D^2
a'\!\cdot\! a'+ \left(1-a'\!\cdot\! R\right)^2\right]R^\mu, \ee
$\hat{\textnormal{\bf D}}^\mu\neq\textnormal{\bf D}^\mu$; note
that $\hat{\textnormal{\bf D}}^\mu\hat{\textnormal{\bf D}}_\mu=-
D^2$, as this was the case for $\textnormal{\bf D}^\mu$ in
(\ref{D.D}). It is clear that $\hat{\textnormal{\bf D}}^\mu\!\!+D
\hat{\tau}^\mu=R^\mu$.

\subsection{Relative three-velocities of reference frames}
\label{s4.3}

Let us now simultaneously consider three distinct reference frames
and denote them as A, B, and C. Between such frames there can be
established quite a few algebraic relations having a clear and
important physical meaning, and it is interesting that these
relations hold equally in general and special relativity. One
defines the relative three-velocity of frame B with respect to
frame A (and measured in A) as a (co)vector $\textnormal{\bf
v}_{\textnormal{\tiny BA}}$ perpendicular to the monad $\tau_{
\textnormal{\tiny A}}$. According to (\ref{v}), \be \label{tau,v}
\tau_{\textnormal{\tiny B}}=(\tau_{ \textnormal{\tiny A}}+
\textnormal{\bf v}_{\textnormal{\tiny BA}})(\tau_{
\textnormal{\tiny A}}\cdot\tau_{\textnormal{\tiny B}})
\textnormal{ and } \textnormal{\bf v}^{\phantom{\textnormal{\tiny
BA}}\mu}_{\textnormal{\tiny BA}}= \frac{\tau^\nu_{
\textnormal{\tiny B}}b^{\phantom{\textnormal{\tiny
A}}\mu}_{\textnormal{\tiny A}\nu}}{\tau_{\textnormal{\tiny
A}}\cdot\tau_{\textnormal{\tiny B}}} \ee (here the relation
$\tau^\mu_{\textnormal{\tiny B}}-\tau^\mu_{ \textnormal{\tiny
A}}(\tau_{\textnormal{\tiny A}}\cdot\tau_{ \textnormal{\tiny
B}})\equiv\tau^\nu_{\textnormal{\tiny
B}}b^{\phantom{\textnormal{\tiny A}}\mu}_{\textnormal{\tiny A}\nu}
$ was used); hence, \be \label{tauA.tauB} \tau_{\textnormal{\tiny
A}}\cdot\tau_{\textnormal{\tiny B}}=\frac{1}{\sqrt{1+
\textnormal{\bf v}_{\textnormal{\tiny BA}}\cdot\textnormal{\bf
v}_{\textnormal{\tiny BA}}}}\equiv\frac{1}{\sqrt{1-
\textnormal{\bf v}_{\textnormal{\tiny BA}}\bullet\textnormal{\bf
v}_{\textnormal{\tiny BA}}}}=\frac{1}{\sqrt{1- \textnormal{\bf
v}_{\textnormal{\tiny BA}}^2}}. \ee It is clear that similar
relations exist for any pair of reference frames whatever when the
respective monads are introduced. We see that there is a symmetry
for squared three-velocities between any pair of frames, in
particular, $\textnormal{\bf v}_{\textnormal{\tiny BA}}^2=
\textnormal{\bf v}_{\textnormal{\tiny AB}}^2$. Since these
three-velocities are described as four-vectors perpendicular to
the respective monads (of the frames corresponding to the frame
subindex of $\tau$ and of $b$), they belong to different (local)
three-spatial sections of spacetime and in general cannot be
directly compared by measurements ones with others without further
projections onto alternative subspaces. The inevitability of such
a situation is quite obvious. Even in the generally used
special-relativistic composition-of-velocities formula for
globally inertial frames in motion along ``same spatial
direction,'' this is in fact also the case which is tacitly
assumed, but frequently not properly understood. Its strict
formulation when these velocities are not mutually ``parallel,''
is however more laborious.

Another useful step in our calculations is to apply same procedure
as in (\ref{tau,v}), but taken with respect to the frames C and A,
then to C and B, and further applying it to the free
$\tau_{\textnormal{\tiny B}}$, thus $\tau_{\textnormal{\tiny
C}}=(\tau_{ \textnormal{\tiny A}}+\textnormal{\bf
v}_{\textnormal{\tiny CA}})(\tau_{\textnormal{\tiny
A}}\cdot\tau_{\textnormal{\tiny C}})=(\tau_{\textnormal{\tiny
B}}+\textnormal{\bf v}_{\textnormal{\tiny
CB}})(\tau_{\textnormal{\tiny B}}\cdot\tau_{\textnormal{\tiny
C}})=[(\tau_{\textnormal{\tiny A}}+\textnormal{\bf
v}_{\textnormal{\tiny BA}})(\tau_{\textnormal{\tiny
A}}\cdot\tau_{\textnormal{\tiny B}}) +\textnormal{\bf
v}_{\textnormal{\tiny CB}}](\tau_{\textnormal{\tiny
B}}\cdot\tau_{\textnormal{\tiny C}})$. When this expression is
multiplied by $b_{\textnormal{\tiny A}}$ under a contraction with
the lower (component) index of this factor, we come to \be
\label{compABC} \textnormal{\bf v}_{\textnormal{\tiny
CA}}^{\phantom{ \textnormal{\tiny CA}}\nu}=\left[\textnormal{\bf
v}_{\textnormal{\tiny BA}}^{\phantom{ \textnormal{\tiny BA}}\nu}
(\tau_{\textnormal{\tiny A}}\cdot\tau_{\textnormal{\tiny B}})+
\textnormal{\bf v}_{\textnormal{\tiny CB}}^{\phantom{
\textnormal{\tiny BA}}\mu}b^{\phantom{\textnormal{\tiny
A}}\nu}_{\textnormal{\tiny
A}\mu}\right]\frac{\tau_{\textnormal{\tiny
B}}\cdot\tau_{\textnormal{\tiny C}}}{\tau_{\textnormal{\tiny
A}}\cdot\tau_{\textnormal{\tiny C}}}. \ee In fact, this is the
local velocities composition formula $\textnormal{A}\rightarrow
\textnormal{B}\rightarrow\textnormal{C}$ for general (not only
inertial) frames in both relativities, special as well as general
one. Here, of course, one has to take into account the relation
(\ref{tauA.tauB}). In this paper we do not consider further
details of the usual composition formula.

Other relations which are worth being mentioned, are the following
ones: those with projections onto the alternative monads, \be
\label{vBAvAB} \textnormal{\bf v}_{\textnormal{\tiny
BA}}^{\phantom{ \textnormal{\tiny BA}}\nu}b^{\phantom{
\textnormal{\tiny B}}\mu}_{\textnormal{\tiny B}\nu}=-
(\tau_{\textnormal{\tiny A}}\cdot\tau_{\textnormal{\tiny B}})
\textnormal{\bf v}_{\textnormal{\tiny AB}}^{\phantom{
\textnormal{\tiny AB}}\mu}\textnormal{ and } \textnormal{\bf
v}_{\textnormal{\tiny AB}}^{\phantom{ \textnormal{\tiny
AB}}\nu}b^{\phantom{ \textnormal{\tiny A}}\mu}_{\textnormal{\tiny
A}\nu}=-(\tau_{\textnormal{\tiny A}}\cdot\tau_{\textnormal{\tiny
B}}) \textnormal{\bf v}_{\textnormal{\tiny BA}}^{\phantom{
\textnormal{\tiny BA}}\mu}; \ee further, due to (\ref{tau,v}) and
(\ref{vBAvAB}), \be \textnormal{\bf v}_{\textnormal{\tiny
AB}}\cdot \textnormal{\bf v}_{\textnormal{\tiny BA}}=-\tau_{
\textnormal{\tiny A}}\cdot\textnormal{\bf v}_{\textnormal{\tiny
AB}}=-(\tau_{\textnormal{\tiny A}}\cdot\tau_{\textnormal{\tiny
B}})\textnormal{\bf v}_{\textnormal{\tiny BA}}^2=(\tau_{
\textnormal{\tiny A}}\cdot\textnormal{\bf v}_{\textnormal{\tiny
AB}})^2/\textnormal{\bf v}_{\textnormal{\tiny AB}}^2 \ee (here the
obvious symmetry $\tau_{ \textnormal{\tiny A}}\cdot\textnormal{\bf
v}_{\textnormal{\tiny AB}}=\tau_{ \textnormal{\tiny
B}}\cdot\textnormal{\bf v}_{\textnormal{\tiny BA}}$ was taken into
account); finally, \be \textnormal{\bf v}_{\textnormal{\tiny AB}}=
-(\tau_{\textnormal{\tiny A}}\cdot\tau_{\textnormal{\tiny B}})
\textnormal{\bf v}_{\textnormal{\tiny BA}}+(\textnormal{\bf
v}_{\textnormal{\tiny AB}}\cdot\tau_{\textnormal{\tiny A}})\tau_{
\textnormal{\tiny A}} \ee (decomposition with respect to the frame
A). Note that $\textnormal{\bf v}_{\textnormal{\tiny AB}}^2:=
\textnormal{\bf v}_{\textnormal{\tiny AB}}\bullet\textnormal{\bf
v}_{\textnormal{\tiny AB}}=-\textnormal{\bf v}_{\textnormal{\tiny
AB}}\cdot\textnormal{\bf v}_{\textnormal{\tiny AB}}>0$.

Let us globally (at any $P$) denote in the LW problem the
reference frame of inertial observer as A,
$\tau_{\textnormal{\tiny A}}^{\phantom{\textnormal{\tiny
A}}\mu}=\delta^\mu_0$, the retarded frame co-moving with the
charge as B, $\tau_{\textnormal{\tiny B}}^{\phantom{
\textnormal{\tiny B}}\mu}=u'^\mu$, and the frame co-moving with
the field and introduced in subsection \ref{s4.2}, as C ($\tau_{
\textnormal{\tiny C}}^{\phantom{ \textnormal{\tiny
C}}\mu}=\hat{\tau}^\mu$). Then, on the one hand, \be (\tau_{
\textnormal{\tiny B}}\cdot\tau_{\textnormal{\tiny C}})=(u'
\cdot\hat{\tau})=1-\frac{1}{2}\left[D^2a'\!\cdot\! a'+ \left( a'\!
\cdot\! R\right)^2\right]. \ee On the other hand, \be
\textnormal{\bf v}_{\textnormal{\tiny CB}}=\frac{\hat{\tau}}{(u'\!
\cdot\!\hat{\tau})}-u'. \ee

Rotation of the frame C takes the (not quite easily deducible)
form \be \label{omegC} \hat{\omega}=-\frac{1- D\,(\dot{a}'\!\cdot
\! R)}{1-a'\!\cdot\! R}\;\textnormal{\bf a}'\!\times\hat{
\textnormal{\bf n}}+D\,\dot{\textnormal{\bf a}}'\!\times\hat{
\textnormal{\bf n}} \ee where 1-form $\dot{a}'
=(da'_\mu/ds')dx^\mu$ describes the retarded third proper-time
derivative of position of the charge in its motion along the
worldline $L$. It is worth giving some hints for the deduction of
(\ref{omegC}): The exterior product of any odd-rank forms $\alpha$
and $\beta$ is skew-symmetric, thus $\alpha\wedge\alpha\equiv 0$.
The vector product (\ref{vecprod}) is applicable to a pair of
arbitrary vectors, thus it automatically projects each of them
onto the three-dimensional subspace orthogonal to the monad. One
now has to apply the definition of rotation (\ref{rot}) to the
monad $\hat{\tau}$. Some simplifications follow immediately. Then
to complete the simplification one has to take into account a
relation following from the form (not directly from the general
definition) of $\hat{\tau}$ (\ref{taucomov}) and $\hat{
\textnormal{\bf D}}$ (\ref{Dhat}):
$$
\hat{ \textnormal{\bf D}}_\mu=D\hat{\tau}_\mu-2D^2a'_\mu-2D(1-a'
\cdot R)u'_\mu+\left[D^2(a'\cdot a')+(1-a'\cdot R)^2\right]R_\mu
$$
(at each subsequent step only very few terms survive). The final
result is (\ref{omegC}) which should be compared with
(\ref{rotu'}).

\section{Concluding remarks} \setcounter{equation}{0} \label{s5}

We tried to give in this paper a self-sufficient consideration of
the LW solution, from its heuristic deduction to an analysis of
important properties of the obtained field. One of these
properties is that of field's motion with respect to a given
reference frame. In fact, one can relate this motion to the monad
describing the frame in which the electromagnetic field does not
propagate (its Poynting vector, the electromagnetic energy flux
density, vanishes in this frame). It is possible to find such a
frame in all cases with the exception of {\it pure null}
electromagnetic fields: in this latter case both electromagnetic
invariants are equal to zero, consequently there remains only an
{\it asymptotic} possibility to transform away the field's motion,
but then {\it it is transformed away always together with the
field itself} (this is precisely the asymptotic limit of the
Doppler effect). This asymptotic situation does not belong to any
admissible reference frame or system of coordinates since such a
frame (or, of one wishes, a system) is a degenerate one and thus
excluded from consideration (whose region of application is an
open one, and the `boundary' is excluded from it, though we can
approach it as `near' as we wish, making the non-zero field as
weak as we choose it to become). In this pure null case (the
definition see in section \ref{s3.1}) the field by itself
exercises lightlike (null) motion, that with the velocity of
light. But then there cannot exist a co-moving (with this field)
reference frame since its four-velocity should coincide with the
monad of the co-moving frame, and the monad vector is timelike by
its definition. (More physical reasons are related to the fact
that the continuous swarm of observers forming, together with
their measuring equipment, a reference frame, and thus being
co-moving with it, should always possess non-zero rest masses,
though, of course, these masses have to be infinitesimal ones to
guarantee the test property of a classical frame of reference. The
non-zero rest mass means a timelike worldline of the corresponding
object, thus the lightlike motion of any reference frame is
physically impossible.) In all other cases concerning
electromagnetic fields' types a co-moving frame is easily
realizable (in this paper we discussed the pure electric and pure
magnetic types, and all impure subcases should be dealt with
according to the method used by Rainich and Wheeler, see
\cite{Whee}).

Another property is also related to propagation, however {\it not
of the field} but {\it of the information} about its sources, thus
this property belongs to {\it the deduction} of the LW field. This
is a rare case when we encounter in a classical physical context
the concept of information usually alien to it. And here
information propagates with the velocity of light in a vacuum.

\renewcommand{\theequation}{\Alph{section}.\arabic{equation}}
\section*{Appendices}
\appendix
\section{Description of reference frames} \label{sA}
\setcounter{equation}{0}

In this paper we use notations and definitions from \cite{Mits96},
see also references therein. A reference frame is understood as
the splitting of general four-dimensional physical quantities into
parts referred to observer's local time direction and the
corresponding local three-dimensional physical subspace orthogonal
to it, however the latter (or both parts) are written as
four-dimensional tensor quantities (of naturally determined ranks)
being orthogonal (or also, if we would wish to emphasize this
geometrically, parallel) to observer's time direction. This
direction is expressed {\it via} the unit vector (or covector, the
distinction should be understandable from the context, frequently
mathematical) $\tau$, the {\it monad}, tangent to the observer's
world line, thus interpreted as the observer's four-velocity at
the event (four-dimensional point) where is located the quantity
(object) under consideration. Thus we speak about a continuous
swarm of observers, a congruence of their world lines without
singularities (the lines do not intersect, and through any event
goes one and only one such line). The monad and the metric tensor
at each event are necessary and sufficient for a complete
description of a reference frame. Of course, this presence of a
swarm of observers, with all their equipment necessary for
measuring of all physical quantities at any event, should not
disturb both usual physical fields and (in general relativity) the
spacetime geometry (the gravitational field). Here we consider
such arbitrary reference frames only in the framework of special
relativity, thus the simplest choice of coordinates is Cartesian
which we use in this paper. In our treatment reference frames are
generally not related to systems of coordinates, and in one and
the same system of coordinates any choice of a reference frame (or
different choices simultaneously) may be used.

To split spacetime tensors into their above-mentioned parts, two
typical projectors are used. A projector is an idempotent, which
means that its repeated action automatically reduces to a single
action of it, and it differs from the metric tensor possessing a
similar (just mentioned) property by the fact that an application
of a projector leads to certain partial loss of information. If we
describe a projector as a $4\times 4$ matrix (really, a rank two
tensor), its determinant should be equal to zero. In more concrete
terms, the matrix rank of a projector should be equal to one when
we speak about a projector onto a single direction (here, $\tau$),
or three when we perform a projection onto the local
three-dimensional physical space orthogonal to $\tau$. Thus in the
first case we can use the projector \be \label{paralproj}
\pi^\mu_\nu=\tau^\mu\tau_\nu \ee and in the second case, \be
\label{bproj} b^\mu_\nu=g^\mu_\nu-\tau^\mu\tau_\nu, \ee hence \be
\label{projprop} \pi^\mu_\lambda\pi^\lambda_\nu=\pi^\mu_\nu, ~ ~
b^\mu_\lambda b^\lambda_\nu=b^\mu_\nu, ~ ~ b^\mu_\nu
\pi^\nu_\lambda=0, ~ ~ b^\mu_\nu\tau^\nu=0. \ee However in the
first case we frequently use a mere interior multiplication (that
is, with a contraction) by $\tau$ since this leads to a
four-dimensionally well defined quantity. It is also clear that
$b^\mu_\nu+\pi^\mu_\nu=g^\mu_\nu$. It is worth being repeated that
the matrices corresponding to (\ref{paralproj}) and (\ref{bproj})
are respectively of ranks one and three.

Traditionally, in the literature one usually finds an implicit
identification of a four-dimensional Cartesian system of
coordinates and the corresponding (``co-moving'') reference frame.
This does not pose any ambiguities, only if different reference
frames are not considered simultaneously on the background of same
system of coordinates, or a non-inertial reference frame is
involved. However it is better to take into account that this
traditional approach represents a tacit admission that the monad
coincides with the unit (timelike) vector along the $t$-axis and
any orthonormal transformation is accompanied with a corresponding
change of the monad. There is also a widespread prejudice that
non-inertial frames cannot be used in or they contradict to the
special theory of relativity, but this is nothing more than a
prejudice. In this paper we consider such frames of non-inertial
observers in two concrete cases, and the monad approach works
perfectly in description of physical situation in these
non-inertial frames. We also use another projector (of rank-two
matrix, that is, realizing projection onto a two-dimensional
subspace) when it simplifies description of the situation, and
there should exist a naturally determined spatial direction which
enables this description.

It is convenient, in the sense of both calculations and adequate
work of physical intuition, to use the vector symbolics of scalar
and vector products denoted as $\bullet$ and $\times$. In fact,
these operations are coincident with those of the
three-dimensional vector algebra, though the objects to which they
are applied are four-dimensional vectors restricted to the
three-dimensional subspace orthogonal to the monad (not always to
the global subspace corresponding in particular to an inertial
frame, but, in rotating frames, changing to the more general local
non-holonomic case: see in the end of this appendix comments
related to the three-dimensional subspaces then having such a
local meaning only). These products are defined as \be
\label{scprod} \textnormal{\bf p}\bullet\textnormal{\bf q}:=-
b_{\mu\nu}p^\mu q^\nu\equiv\ast[(\tau \wedge p)\wedge\ast(\tau
\wedge q)] \ee and \be \label{vecprod} \textnormal{\bf p}\times
\textnormal{\bf q}=\ast(p\wedge\tau \wedge q). \ee We use here the
Cartan exterior forms notations such as the wedge pro\-duct
$\wedge$, the Hodge star operation $\ast$ (the dual conjugation of
a $p$-form, not necessarily of a 2-form = skew-symmetric rank-two
tensor), and, later, the exterior differential $d$, see for
details and references \cite{Mits96}.

In Cartesian coordinates, due to the spacetime signature
$(+,-,-,-)$, the monad of the frame co-moving with these
coordinates is $\tau^\mu=\delta^\mu_0, ~ \tau_\mu=\delta_\mu^0$.
Thus (\ref{vecprod}) becomes $(\textnormal{{\bf p}}\times
\textnormal{{\bf q}})^i=\epsilon_{ijk}p^jq^k$. The (co)vectors
lying in the three-dimensional subspace of a reference frame are
usually written as four-dimensional ones, but in some important
cases we put them in boldface printing (as {\bf E} and {\bf B} for
electric and magnetic vectors). Then {\bf E}$^2\equiv
\textnormal{{\bf E}}\bullet\textnormal{{\bf E}}=-b_{\mu\nu}
\textnormal{{\bf E}}^\mu\textnormal{{\bf E}}^\nu$, {\it etc}.

The three-dimensional velocity {\bf v} (described as a four-vector
$\perp\tau$) of a pointlike particle from the viewpoint of
reference frame corresponding to the monad $\tau$, is determined
{\it via} the splitting of its four-velocity $u^\mu=dx^\mu/ds$,
\be \label{v} u=(\tau\cdot u)(\tau+\textnormal{{\bf
v}}),\textnormal{ or equivalently } \textnormal{{\bf
v}}^\mu\!=b^\mu_\nu \frac{dx^\nu}{\tau_\alpha dx^\alpha} \ee where
$\tau_\alpha dx^\alpha/ds=(1-v^2)^{-1/2}$; {\it cf.} also
(\ref{tauA.tauB}) and the corresponding remarks. This is, of
course, an exclusion in the general method of projecting vector
and tensor quantities. Another exclusion is the relation between
the four-dimensional acceleration and its usual three-dimensional
counterpart which is applied in making an easier comparison with
the Landau--Lifshitz treatment of the LW field \cite{LanLif}. It
is now convenient to write the corresponding relations in the
(local) three-dimensional subspace notations. The relativistic
acceleration four-vector then is
$$
a'^\mu=\frac{du'^\mu}{ds'}=\frac{1}{\sqrt{1-\textnormal{\bf
v}'^2}} \left[\frac{d}{dt'}\left(\frac{1}{\sqrt{1-\textnormal{\bf
v}'^2}} \right)\right](1,\textnormal{\bf
v}')+\frac{1}{1-\textnormal{\bf v}^2}\left(0,\dot{\textnormal{\bf
v}'} \right),
$$
and the orthogonality of $a'$ and $u'$, \be \label{ua-orth}u'\cdot
a'= \frac{d}{dt'}\left(\frac{1}{\sqrt{1-\textnormal{\bf
v}'^2}}\right)-\frac{1}{\left(1-\textnormal{\bf
v}'^2\right)^{3/2}}\textnormal{\bf
v}'\!\!\bullet\!\dot{\textnormal{\bf v}'}=0, \ee finally yields a
simpler relation between the four- and three-acceleration \be
\label{a4-3} a'^\mu=\frac{\textnormal{\bf
v}'\!\!\bullet\!\dot{\textnormal{\bf v}'}}{\left(1-\textnormal{\bf
v}'^2\right)^2}(1,\textnormal{\bf v}')+ \frac{1}{1-\textnormal{\bf
v}'^2}\left(0,\dot{\textnormal{\bf v}'}\right). \ee

Rotation of a reference frame is defined as \be \label{rot}
\omega=\ast(\tau \wedge d\tau)\equiv 2\ast(\tau\wedge A), ~ ~
A=\frac{1}{2} A_{\mu\nu}dx^\mu\wedge dx^\nu, \ee while in
Cartesian coordinates and with $\tau$ describing a non-inertial
frame, $A$ (not the electromagnetic four-potential 1-form, but the
rotation 2-form) is the skew term in the natural decomposition of
gradient of the monad, \be \label{gradmon} \tau_{\mu,\nu}=\tau_\nu
G_\mu+A_{\nu\mu}+D_{\nu\mu}, ~ ~ A_{\mu\nu}=A_{[\mu\nu]}, ~ ~
D_{\mu\nu}=D_{(\mu\nu)}, \ee $G$ being acceleration of the
reference frame and $D$, the frame's symmetric rate-of-strain
tensor; $G$, $A$, and $D$ belong to the above-mentioned
three-dimensional (local) subspace. Of course, all these
quantities become equal to zero in any inertial frame globally.
When $A\neq 0$ (equivalent to $\omega\neq 0$), the
three-dimensional subspace orthogonal to $\tau$ is non-holonom,
that is, there only exists an overall distribution of elements of
the corresponding (now non-holonom) hypersurface, but these
elements do not fit together to form a global spatial hypersurface
in the proper (holonom) sense, see \cite{Mits96}, the fact well
known in geometry of congruences (here we are dealing with the
$\tau$-congruence).

\section{Electromagnetic fields in arbitrary
reference frames} \label{sB} \setcounter{equation}{0}

Let us now apply the definitions given in appendix \ref{sA} to the
electromagnetic field and related quantities. The field tensor
$F_{\alpha\beta}$ which also can be written as a 2-form \be
\label{F2form} F=\frac{1}{2}F_{\mu\nu}dx^\mu\wedge dx^\nu, \ee
splits into two four-dimensional vectors, electric \be \label{elE}
\textnormal{{\bf E}}_\mu=F_{\mu\nu} \tau^\nu ~ ~
\Longleftrightarrow ~ ~ \textnormal{{\bf E}}= \ast(\tau \wedge\ast
F) \ee and magnetic \be \label{magB} \textnormal{{\bf
B}}_\mu=-F\!\!\stackrel{\textnormal{\small$\ast$}
}{\textnormal{\scriptsize$\mu\nu$}}\!\tau^\nu ~ ~
\Longleftrightarrow ~ ~ \textnormal{{\bf B}}=\ast(\tau\wedge F),
\ee both $\perp\tau$, see also (\ref{Fdual}); 2-form $F:=
\frac{1}{2}F_{\mu\nu}dx^\mu \wedge dx^\nu$. This splitting follows
from an observation that the Lorentz force can be expressed as \be
\label{LorForce} (\textnormal{{\bf E}}+\textnormal{{\bf v}}\times
\textnormal{{\bf
B}})_\alpha=F_{\mu\nu}\left(\tau^\nu+\textnormal{{\bf
v}}^\nu\right) b^\mu_\alpha. \ee In Cartesian coordinates (and
with the corresponding inertial monad) we have the same relations
as for usual contravariant three-vectors: \be \label{EB}
\textnormal{{\bf E}}^i=F_{i0}=-F^{i0}, ~ ~ \textnormal{{\bf
B}}^i=-\frac{1}{2} \epsilon_{ijk}F_{jk}=-\frac{1}{2}
\epsilon_{ijk}F^{jk}, \ee thus \be \label{Fij}F_{ij}=F^{ij}=-
\epsilon_{ijk}\textnormal{{\bf B}}^k. \ee

The electromagnetic stress-energy tensor is \be \label{Tmunu}
{T_{\textnormal{{\scriptsize em}}}}^\nu_\mu=\frac{1}{4\pi}\left(
\frac{1}{4}F_{\kappa\lambda}F^{\kappa\lambda} \delta^\nu_\mu-
F_{\mu\lambda}F^{\nu\lambda}\right) \ee (in Gaussian units). Its
deduction is most simple when one considers Max\-well's equations
in tensor form in a vacuum and without sources. Its (single)
contraction with arbitrary monad includes the Poynting vector in
that frame, \be \label{Ttau} {T_{\textnormal{{\scriptsize
em}}}}^\nu_\mu\tau_\nu= \frac{1}{8\pi}
\left[\left(\textnormal{{\bf E}}^2+\textnormal{{\bf
B}}^2\right)\tau_\mu+2(\textnormal{{\bf E}}\times\textnormal{{\bf
B}})_\mu\right], \ee and the squared expression is
$$
{T_{\textnormal{{\scriptsize em}}}}^\nu_\mu
{T_{\textnormal{{\scriptsize em}}}}^\mu_\xi\tau_\nu\tau^\xi=
\frac{1}{(8\pi)^2} \left[\left(\textnormal{{\bf
E}}^2+\textnormal{{\bf B}}^2\right)^2-4(\textnormal{{\bf E}}\times
\textnormal{{\bf B}})^2\right]
$$
\be \label{Ttausq} \equiv\frac{1}{(8\pi)^2}\left[
\left(\textnormal{{\bf B}}^2-\textnormal{{\bf
E}}^2\right)^2+4(\textnormal{{\bf E}} \bullet\textnormal{{\bf
B}})^2\right]=\frac{1}{(16\pi)^2} \left({I_1}^2+{I_2}^2\right) \ee
(it is interesting that this expression is not only a scalar under
transformations of coordinates, but it is also independent of the
choice of reference frame: the right-hand side does not involve
any mention of the monad at all). For the LW field [due to
(\ref{I1})] this takes a very concise form, \be \label{LWTtausq}
{T_{\textnormal{{\scriptsize em}}}}^\nu_\mu
{T_{\textnormal{{\scriptsize em}}}}^\mu_\xi\tau_\nu\tau^\xi=
\left(\frac{Q^2}{8\pi D^4}\right)^2. \ee

\end{document}